\documentclass[%
 reprint,
 superscriptaddress,
showpacs,%preprintnumbers,
 amsmath,amssymb,
 aps,
]{revtex4-1}

\usepackage{graphicx}% Include figure files
\usepackage{dcolumn}% Align table columns on decimal point
\usepackage{bm}% bold math
\newcommand{\tr}{\operatorname{tr}}
\newcommand{\rmd}{\mathrm{d}}
\newcommand{\nn}{\nonumber}
\newcommand{\cC}{\mathcal C}

\newcommand{\cM}{\mathcal M}
\newcommand{\cP}{\mathcal P}
\newcommand{\cQ}{\mathcal Q}
\newcommand{\cR}{\mathcal R}
\newcommand{\cS}{\mathcal S}

\begin{document}

%\preprint{APS/123-QED}

\title{\hspace*{-0.7cm}Branes in Extended Spacetime: Brane Worldvolume Theory Based on Duality Symmetry}

\author{Yuho Sakatani}
\email{yuho@koto.kpu-m.ac.jp}
\affiliation{%
Department of Physics, Kyoto Prefectural University of Medicine, Kyoto 606-0823, Japan
}%
\affiliation{
B.W.~Lee Center for Fields, Gravity \& Strings, CTPU, Institute for Basic Sciences, Daejeon 34047, Korea
}%
\author{Shozo Uehara}%
 \email{uehara@koto.kpu-m.ac.jp}
\affiliation{%
Department of Physics, Kyoto Prefectural University of Medicine, Kyoto 606-0823, Japan
}%

\begin{abstract}
We propose a novel approach to the brane worldvolume theory based on the geometry of extended field theories: double field theory and exceptional field theory. We demonstrate the effectiveness of this approach by showing that one can reproduce the conventional bosonic string and membrane actions, and the M5-brane action in the weak-field approximation. At a glance, the proposed 5-brane action without approximation looks different from the known M5-brane actions, but it is consistent with the known nonlinear self-duality relation, and it may provide a new formulation of a single M5-brane action. Actions for exotic branes are also discussed. 
\end{abstract}

%\pacs{
%11.25.Yb,%M theory
%11.25.-w,%Strings and branes
%04.65.+e,%Supergravity
%}% PACS, the Physics and Astronomy
\maketitle

%\tableofcontents
\setcounter{equation}{0}

%\section{Introduction}
{{\it Introduction}.---}String theory or M-theory has a duality symmetry when compactified on a torus. 
Recently, it has become possible to formulate the effective supergravity actions in a manifestly duality invariant manner due to the developments in extended field theories, the double field theory (DFT) \cite{Siegel:1993xq,Siegel:1993th,Siegel:1993bj,Hull:2009mi,Hohm:2010pp} and the exceptional field theory (EFT) \cite{West:2000ga,West:2001as,Hillmann:2009pp,Berman:2010is,West:2012qz,Hohm:2013pua}. 
These extended field theories can describe novel geometries, such as non-geometric backgrounds that naturally appear in string or M-theory, and further interesting geometries beyond the scope of the conventional supergravity are under investigation. 

In worldvolume theories, duality covariant equations of motion and duality invariant actions were developed in \cite{Duff:1989tf,Tseytlin:1990nb,Tseytlin:1990va,Hull:2004in,Hull:2006va,Copland:2011wx,Lee:2013hma} for a string, and in \cite{Duff:1990hn,Percacci:1994aa,Lukas:1997jk,Bengtsson:2004nj,West:2004iz,Berman:2010is,Hatsuda:2012vm,Hatsuda:2013dya,Duff:2015jka,Hu:2016vym} for M-theory branes. 
Specifically, the double sigma model provides a $T$-duality invariant string action, and with which we can study the string theory in various nontrivial backgrounds (see \cite{Ko:2015rha} for an application). 
In contrast, a similar formulation for M-theory branes that respects the duality symmetries is still not well developed. 

In this Letter, we propose a general construction of brane actions that respects the geometry on extended spacetimes: the doubled spacetime in DFT and the exceptional spacetime in EFT. 
Namely, we construct brane actions that are invariant under the generalized diffeomorphisms in the extended spacetimes and also a subgroup of the duality symmetry, which we explain later. 
For that purpose, we utilize the untwisting procedure developed in \cite{Hull:2014mxa,Rey:2015mba,Chaemjumrus:2015vap}. 
Our untwisting matrix consists of dynamical worldvolume gauge fields, and they, together with the conventional embedding functions $X^i$, describe the embedding into the extended spacetime. 
A similar idea is proposed in \cite{Asakawa:2012px}, where the effective action for a D-brane was formulated by treating the scalar fields $X^i$ and gauge fields $A_a$ on an equal footing. 
There, $X^i$ and $A_a$ describe the fluctuations in the physical and the dual directions, respectively, and the dynamics of a D-brane in the doubled spacetime is described by the pair, $(X^i,\,A_a)$. 
Our formulation extends their idea to a worldvolume theory in an arbitrary extended spacetime. 

%\section{Geometry on extended spacetimes}
{{\it Geometry on extended spacetimes}.---}Let us begin with a brief explanation of the geometry on extended spacetimes (see \cite{Rey:2015mba} for more details). 
We consider a certain extended spacetime as the target space, which has the local coordinates $x^I$ and the generalized metric $\cM_{IJ}(x)$. 
We decompose $x^I$ into the coordinates $x^i$ on the physical $d$-torus and the dual coordinates $y_M$ supposing that all fields and gauge parameters have only $x^i$ dependence. 
In DFT and EFT, $\cM_{IJ}$ always has the factorized form, $\cM_{IJ} = \hat{\cM}_{KL}\, L^K{}_I\,L^L{}_J$, and $\hat{\cM}_{KL}$ is a block-diagonal matrix including only the metric $G_{ij}$ on the physical $d$-torus, while $L_I{}^J$ consists of various $(q+1)$-form potentials in the supergravity. 
For example, in doubled \cite{Duff:1989tf,Siegel:1993th} and $E_{6(6)}$-exceptional spacetime \cite{West:2003fc,Berman:2011jh}, $x^I$ and the set of $(q+1)$-form potentials are given by
\begin{align}
 &\text{DFT: }\ (x^I)=\bigl(x^i,\,\tilde{x}_i\bigr)\,,\quad \{B_{ij}\}\,,
\nn\\
 &\text{EFT: }\ (x^I)=\bigl(x^i,\,y_{i_1i_2},\,y_{i_1\cdots i_5}\bigr)\,,\quad \{C_{i_1i_2i_3},\,C_{i_1\cdots i_6}\}\,.
\nn
\end{align}
Note that a $(q+1)$-form potential exists for each dual coordinate, with $q$ totally antisymmetric indices. 
From the perspective developed in \cite{Rey:2015mba}, an extended spacetime is foliated by a family of physical $d$-tori, and, because of the section condition in extended field theories, the foliation is uniform in the orthogonal dual directions. 
The shape of the foliation can be specified by a set of closed $(q+1)$-form fields $\{c_{q+1}(x)\}$, which are also in one-to-one correspondence with dual coordinates with $q$ indices and transform in the same manner as the potentials $\{C_{q+1}\}$ under generalized diffeomorphisms \cite{Rey:2015mba}. 
A physical point in the extended space can then be specified by the foliation $\{c_{q+1}(x)\}$ and coordinates $x^i$ on the physical $d$-tori. 

Under a generalized diffeomorphism that maps the ``generalized point'' $(x^i,\,\{c_{q+1}(x)\})$ into $(x'^i,\,\{c'_{q+1}(x')\})$, the transformation law of a generalized tensor $T^{I_1\cdots I_r}_{J_1\cdots J_s}(x)$ was found to be \cite{Rey:2015mba}
\begin{align}
\begin{split}
 T'^{I_1\cdots I_r}_{J_1\cdots J_s}(x') &= \cS^{I_1}{}_{K_1}\cdots (\cS^{-1})^{L_1}{}_{J_1}\cdots T^{K_1\cdots K_r}_{L_1\cdots L_s}(x)\,,
\\
 \cS^I{}_J&\equiv \bigl(E'^{-1}\bigr)^I{}_K(x')\,\cR^K{}_L\,E^L{}_J(x)\,. 
\end{split}
\end{align}
Here, we defined a matrix $\cR^I{}_J$ by replacing the matrix $G_{ij}$ contained in $\hat{\cM}_{IJ}$ with the Jacobian matrix, $\partial x'^i/\partial x^j$, and defined a matrix $E^I{}_J$ by replacing $C_{q+1}(x)$ in $L^I{}_J$ with the closed form, $c_{q+1}(x)$. 

%\section{Worldvolume theory}
{{\it Worldvolume theory}.---}Now we consider a $p$-brane with the intrinsic metric $\gamma_{\alpha\beta}$ and the local coordinates $\sigma^\alpha$. 
If the worldvolume $\Sigma$ is on the physical $d$-torus, its position is parametrized only by the conventional embedding functions $X^i(\sigma)$. 
In general, the $p$-brane can also fluctuate along the dual directions, and the fluctuation along each dual direction, $y_{i_1\cdots i_q}$, can be parametrized by a locally closed-form field on the worldvolume, $F_{q+1}(\sigma) = \rmd A_{q}(\sigma)$ (note that a $q$-form $A_q(\sigma)$ on the worldvolume vanishes for $q>p+1$). 
Under generalized diffeomorphisms, $\{F_{q+1}(\sigma)\}$ transforms in the same manner as the pullback of $\{c_{q+1}(x)\}$ for the embedding map $X^i(\sigma)$. 
Note that $F_{q+1}(\sigma)$ is different from the pullback of $c_{q+1}(x)$; $c_{q+1}(x)$ specifies the embedding of the physical $d$-torus while $F_{q+1}(\sigma)$ specifies that of the $p$-brane. 

We further introduce auxiliary 1-form fields $\cP_M(\sigma)$ and define the 1-form, $(\cP^I)\equiv (\rmd X^i,\cP_M)$, that transforms as a generalized vector. 
For example, $(\cP^I)=(\rmd X^i,\,\cP_i)$ in DFT and $(\cP^I)=\bigl(\rmd X^i,\, \cP_{i_1i_2},\,\cP_{i_1\cdots i_5}\bigr)$ in EFT. 
The fundamental fields are summarized as follows:
\begin{align}
 \bigl\{\,\gamma_{\alpha\beta}(\sigma),\,\ X^i(\sigma),\,\  A_q(\sigma),\,\  \cP_M(\sigma)\,\bigr\}\,. 
\end{align}

Our action for a single $p$-brane is given by
\begin{align}
 S = \frac{1}{p+1}\biggl[\,\frac{1}{2}\int_\Sigma \cM_{IJ}(X)\,\cP^I\wedge *\,\cP^J - \int_\Sigma \Omega_{p+1}\,\biggr] \,. 
\end{align}
The first term is manifestly invariant under duality transformations and generalized diffeomorphisms. 
The second term $\Omega_{p+1}$ is roughly given by $\Omega_{p+1}\sim \cP_{i_1\cdots i_p}\wedge\rmd X^{i_1\cdots i_p}$ with $\rmd X^{i_1\cdots i_p}\equiv\rmd X^{i_1}\wedge\cdots\wedge\rmd X^{i_p}/\sqrt{p!}$\,. 
In order to describe the fluctuation of the $p$-brane, we include $F_{q+1}(\sigma)$ in the definition of $\Omega_{p+1}$, such that $\Omega_{p+1}$ becomes invariant under generalized diffeomorphisms. 
If we define the untwisted vector, $\widehat{\cP}^I\equiv L^I{}_J\,\cP^J$, it transforms as $\widehat{\cP}'^I=\cR^I{}_J\,\widehat{\cP}^J$ under generalized diffeomorphisms \cite{Hull:2014mxa}. 
Since $\{F_{q+1}(\sigma)\}$ transforms in the same manner as the pullback of $\{C_{q+1}(x)\}$, if we define $\check{\cP}^I$ by replacing $C_{q+1}$ in $\widehat{\cP}^I$ with $F_{q+1}$, it also transforms as $\check{\cP}'^I=\cR^I{}_J\,\check{\cP}^J$. 
Since $\check{\cP}_{i_1\cdots i_p}$ transforms as $\check{\cP}'_{i_1\cdots i_p} = \cR_{i_1\cdots i_p}{}^{j_1\cdots j_p}\,\check{\cP}_{j_1\cdots j_p}$ and $\rmd X^{j_1\cdots j_p}$ transforms as a $p$-vector, $\rmd X'^{i_1\cdots i_p} = (\cR^{-1})_{j_1\cdots j_p}{}^{i_1\cdots i_p}\,\rmd X^{j_1\cdots j_p}$, $\Omega_{p+1}\equiv \check{\cP}_{i_1\cdots i_p}\wedge\rmd X^{i_1\cdots i_p}$ is invariant under generalized diffeomorphisms. 

Unlike the first term in the action that is invariant under the whole duality symmetry, the second term is invariant only under a subgroup. 
In fact, in order to meet the consistency condition (i.e., section condition) of DFT or EFT, we suppose that gauge parameters are independent of the dual coordinates. 
Then, generalized diffeomorphisms reduce to the gauge transformations of the supergravity, called the geometric subgroup: i.e., diffeomorphisms on the torus and the gauge transformations of the gauge potentials \cite{Hull:2009zb}. 
By construction, the second term is invariant under the restricted duality transformations contained in the geometric subgroup, but not under the whole duality transformations. 
This is reasonable since, under a generic transformation, a $p$-brane is transformed to another brane and $\Omega_{p+1}$ should be changed. 

A consistent result was obtained in \cite{Percacci:1994aa,Duff:2015jka,Hu:2016vym}; the duality symmetry in a brane worldvolume theory is realized as a symmetry that mixes the Bianchi identities and the equations of motion \cite{Duff:1989tf,Duff:1990hn,Percacci:1994aa,Hull:2004in}, but, as was found in \cite{Percacci:1994aa,Duff:2015jka,Hu:2016vym}, only the subgroup (i.e., the geometric subgroup) of the $U$-duality symmetry is consistently realized (classically). 
Specifically, in the case of the $\mathrm{SL}(5)$ $U$-duality symmetry, the duality transformations are parametrized by $a^i{}_j$ and $b_{ijk}$ (i.e., global $\mathrm{GL}(4)$ transformations and constant shifts in $C_{ijk}$) together with $c^{ijk}$ (the non-geometric $\Omega$-shift \cite{Malek:2012pw}). 
Only the geometric subgroup satisfying $c^{ijk}=0$ was shown to be the symmetry of the equations of motion \cite{Duff:2015jka}, and our action is also invariant only under the same subgroup. 

There is also a more ambitious attempt to construct the $U$-duality invariant action \cite{Bengtsson:2004nj}. 
It is, however, quite challenging since, for that purpose, we need to treat all branes with different dimensionality on an equal footing. 

%\section{String in doubled spacetime}
{{\it String in doubled spacetime}.---}As the simplest application of our formulation, let us consider a fundamental string in a doubled spacetime. 
In the doubled spacetime, the generalized metric is parametrized as
\begin{align}
 (\cM_{IJ})=\begin{pmatrix} \mathbf{1} & B \cr \mathbf{0} & \mathbf{1}
 \end{pmatrix}\begin{pmatrix} G & \mathbf{0} \cr \mathbf{0} & G^{-1}
 \end{pmatrix}\begin{pmatrix} \mathbf{1} & \mathbf{0} \cr -B & \mathbf{1}
 \end{pmatrix} \,,
\label{eq:DFT-gen-met}
\end{align}
and the 1-form fields are given by $(\cP^I)=(\rmd X^i,\,\cP_i)$\,. 
The 2-form $\Omega_2$ is then given by
\begin{align}
 \Omega_2=\bigl(\cP_i - F_{ij}\,\rmd X^j\bigr)\wedge \rmd X^i = \cP_i\wedge\rmd X^i  + 2\, F_2 \,,
\end{align}
and the action becomes
\begin{align}
 \!S =\! \int_\Sigma \biggl[\,\frac{1}{4}\, \cM_{IJ}(X)\,\cP^I\wedge *\,\cP^J - \frac{1}{2}\,\cP_i\wedge\rmd X^i - F_2\biggr] \,.
\label{eq:our-string}
\end{align}
The equation of motion for $\cP_i$ gives
\begin{align}
 \cP_i = B_{ij}\,\rmd X^j + G_{ij}* \rmd X^j \,, 
\end{align}
and substituting this into the action, we obtain an equivalent action of the well-known form
\begin{align}
 S = \frac{1}{2}\int_\Sigma G_{ij}\,\rmd X^i\wedge * \rmd X^j + \int_\Sigma B_2 - \int_{\partial\Sigma} A_1 \,, 
\end{align}
where $B_2\equiv (1/2)\,B_{ij}\,\rmd X^i\wedge\rmd X^j$ and the worldsheet is supposed to have the boundary $\partial\Sigma$. 
Note that our action is equivalent to the conventional sigma model action in arbitrary curved backgrounds. 

Let us compare our theory with the double sigma model. 
Apart from the total-derivative term, $\int_\Sigma F_2$\,, our action has a similar structure to that of Hull's double sigma model \cite{Hull:2004in,Hull:2006va} or a similar model by Lee and Park \cite{Lee:2013hma}. 
Indeed, if we replace our $\cP^I$ with $\cP^I\equiv \rmd \mathbb{X}^I +C^I$, $(\mathbb{X}^I)\equiv(X^i,\,\tilde{X}_i)$, and $(C^I)\equiv(0,C_i)$, the action \eqref{eq:our-string} becomes Hull's action, in a version ``doubled everything,''
\begin{align}
 S =\! \int_\Sigma \Bigl[\,\frac{1}{4}\,\cM_{IJ}\,\cP^I\wedge * \cP^J - \frac{1}{2}\bigl(\rmd \tilde{X}_i+C_i\bigr)\wedge \rmd X^i \Bigr] \,. 
\end{align}
Lee and Park's action is also obtained by adding the total-derivative term, $(1/2)\int_\Sigma \rmd \tilde{X}_i\wedge \rmd X^i$\,. 
If we consider a constant background, our equation of motion for $X^i$ gives $\rmd \cP_i=0$ and we can find $\tilde{X}_i$ as a solution of $\cP_i=\rmd \tilde{X}_i$. 
However, in general backgrounds, we may not solve for $\tilde{X}_i$ and it will be a key difference of our approach from the double sigma model. 
Furthermore, in our approach, because of the introduction of $F_2=\rmd A_1$---which parametrizes the fluctuations of a string along the dual directions---the boundary term for an open string is reproduced correctly. 

%\section{M-branes in exceptional spacetime}
{{\it M-branes in exceptional spacetime}.---}We now consider branes in the $E_{d(d)}$-exceptional spacetime with the generalized metric $\cM_{IJ}(X)$. 
For the notational simplicity, we consider the $27$-dimensional $E_{6(6)}$-exceptional spacetime, and we ignore the dynamics of branes in the uncompactified five dimensions. 
The 1-form fields $\cP^I$ in this case are given by $(\cP^I)=\bigl(\rmd X^i,\, \cP_{i_1i_2},\,\cP_{i_1\cdots i_5} \bigr)$\,. 
In order to consider the time evolution, we choose the time coordinate as one of the compactified six-torus, as was done in the $\mathrm{SL}(5)$ case \cite{Duff:1990hn,Berman:2010is} (or in DFT or EFT \cite{Berkeley:2014nza,Berman:2014jsa}). 
The case of $E_{d(d)}$ EFT with a smaller $d$ can be considered simply by restricting the range of the index $i$. 
The $E_{d(d)}$ case with $d=7,\,8$ can also be considered by using the generalized metric obtained in \cite{Berman:2011jh,Godazgar:2013rja}. 

The generalized metric in $E_{6(6)}$ EFT has the form \cite{Berman:2011jh}
\begin{align}
\begin{split}
 &\cM_{IJ} = \hat{\cM}_{KL}\, L^K{}_I\,L^L{}_J \,,
\\
 &\bigl(\hat{\cM}_{IJ}\bigr) \equiv{\scriptsize\begin{pmatrix}
 G_{ij} & \mathbf{0} & \mathbf{0} \\
 \mathbf{0} & G^{i_1i_2,j_1j_2} & \mathbf{0} \\
 \mathbf{0} & \mathbf{0} & G^{i_1\cdots i_5,j_1\cdots j_5}
 \end{pmatrix}} \,,
\\
 &\bigl(L^I{}_J\bigr) \equiv{\scriptsize\begin{pmatrix}
 \delta^i_j & \mathbf{0} & \mathbf{0} \\
 \frac{1}{\sqrt{2}}\, C_{i_1i_2 j} & \delta_{i_1i_2}^{j_1j_2} & \mathbf{0} \\
 L_{i_1\cdots i_5,j}  & \frac{10\sqrt{2}}{\sqrt{5!}}\, \delta^{j_1j_2}_{[i_1i_2} C_{i_3i_4i_5]} & \delta_{i_1\cdots i_5}^{j_1\cdots j_5} 
 \end{pmatrix}} \,,
\\
 &L_{i_1\cdots i_5,j}\equiv -\frac{1}{\sqrt{5!}}\, (C_{i_1\cdots i_5 j}-5\, C_{[i_1i_2i_3}C_{i_4i_5] j}) \,,
\end{split}
\end{align}
where $\delta_{i_1\cdots i_q}^{j_1\cdots j_q}\equiv \delta_{[i_1}^{j_1}\cdots \delta_{i_q]}^{j_q}$ and $G^{i_1\cdots i_q,j_1\cdots j_q}\equiv \delta_{k_1\cdots k_q}^{i_1\cdots i_q}\,G^{k_1j_1}\cdots G^{k_qj_q}$, and we also define $G_{i_1\cdots i_q,j_1\cdots j_q}\equiv \delta^{k_1\cdots k_q}_{i_1\cdots i_q}\,G_{k_1j_1}\cdots G_{k_qj_q}$\,. 
We can calculate the $(p+1)$-form $\Omega_{p+1}$ for a $2$-brane and a $5$-brane as follows:
\begin{align}
\begin{split}
 \Omega_3 &\equiv \cP_{i_1i_2}\wedge \rmd X^{i_1i_2} + 3\,F_3 \,,
\\
 \Omega_6 &\equiv \cP_{i_1\cdots i_5}\wedge \rmd X^{i_1\cdots i_5} + \cP_{ij}\wedge \rmd X^{ij} \wedge F_3 +6\,F_6\,. 
\end{split}
\end{align}

%\section{M2-brane action}
{\underline{\it M2-brane action:}}\quad
Our bosonic action for a single membrane becomes
\begin{align}
 S=\frac{1}{3}\int_\Sigma\biggl(\frac{1}{2}\,\cM_{IJ}\,\cP^I\wedge *\cP^J -\Omega_3 \biggr)\,. 
\label{eq:M2-action}
\end{align}
The equation of motion for $\cP_{i_1\cdots i_5}$ simply gives an algebraic relation and, using that, we obtain the action
\begin{align}
\begin{split}
 S&=\int_\Sigma\Bigl[\frac{1}{6}\,G_{ij}\,\rmd X^i \wedge *\rmd X^j - \frac{1}{3}\, \cP_{i_1i_2}\wedge \rmd X^{i_1 i_2}
\\
 &\qquad + \frac{1}{6}\,G^{i_1i_2,j_1j_2}\,\Bigl(\cP_{i_1i_2}+\frac{1}{\sqrt{2}}\,C_{i_1i_2 k}\,\rmd X^k\Bigr) 
\\
 &\qquad\qquad \wedge *\Bigl(\cP_{j_1j_2}+\frac{1}{\sqrt{2}}\,C_{j_1j_2 l}\,\rmd X^l\Bigr) \Bigr]
  - \int_\Sigma F_3 \,. 
\end{split}
\end{align}
The equation of motion for $\cP_{i_1i_2}$ gives
\begin{align}
 \cP_{i_1i_2} = -\frac{1}{\sqrt{2}}\, C_{i_1i_2 j}\, \rmd X^j 
 - G_{i_1i_2, j_1j_2}\, *\rmd X^{j_1j_2} \,,
\end{align}
and, substituting this further into the action, we obtain
\begin{align}
 S &= \frac{1}{6}\int_\Sigma G_{ij}\,\rmd X^i\wedge *\rmd X^j  
 + \int_\Sigma \bigl(C_3-F_3\bigr)
\nn\\
 &\quad -\frac{1}{6}\int_\Sigma G_{i_1i_2,j_1j_2} * \rmd X^{i_1i_2} \wedge *\,(*\,\rmd X^{j_1j_2}) \,.
\end{align}
This is similar to the action obtained from the non-topological Nambu sigma model (see (4.4) in \cite{Schupp:2012nq}). 
If we define the induced metric by $h_{\alpha\beta}\equiv G_{ij}\,\partial_\alpha X^i\,\partial_\beta X^j$, the equation of motion for $\gamma_{\alpha\beta}$ can be written as
\begin{align}
 h_{\alpha\beta}= \frac{\det h}{\det\gamma}\, (\gamma h^{-1}\gamma)_{\alpha\beta} \,,
\end{align}
and it shows $\gamma_{\alpha\beta}=h_{\alpha\beta}$ if it is satisfied at the initial time. 
Using this, we obtain the conventional membrane action \cite{Bergshoeff:1987cm} including the boundary term \cite{Brax:1997ka}:
\begin{align}
 S =- \int_\Sigma \rmd^3\sigma \sqrt{-h} + \int_\Sigma C_3 - \int_{\partial\Sigma} A_2 \,.
\end{align}
In constant backgrounds, the equation of motion for $X^i$ gives $\rmd \cP_{ij}=0=\rmd \cP_{i_1\cdots i_5}$, and we can calculate the dual coordinates via $\cP_{ij}=\rmd Y_{ij}$ and $\cP_{i_1\cdots i_5}=\rmd Y_{i_1\cdots i_5}$ once a classical solution is found. 
Combining these, we find all coordinates $\mathbb{X}^I$ as we do in the case of the double sigma model.

%\section{M5-brane action}
{\underline{\it M5-brane action:}}\quad
The bosonic action for a 5-brane in the $E_{6(6)}$-exceptional space becomes
\begin{align}
 S= \frac{1}{6}\int_\Sigma\biggl(\frac{1}{2}\,\cM_{IJ}\,\cP^I\wedge *\cP^J - \Omega_6\biggr)\,. 
\end{align}
Similar to the membrane case, using the equations of motion for $\cP_{i_1\cdots i_5}$ and $\cP_{i_1i_2}$, we can eliminate these auxiliary fields to obtain the equivalent action,
\begin{align}
 S&= -\frac{1}{12}\int_\Sigma\rmd^6\sigma \Bigl[\sqrt{-\gamma}\,\gamma^{\alpha\beta}\,h_{\alpha\beta} - \frac{\det h}{\sqrt{-\gamma}}\,\theta^\alpha{}_\beta\,(h^{-1}\gamma)^\beta{}_\alpha \Bigr] 
\nn\\
 &\quad + \int_\Sigma \Bigl(C_6-\frac{1}{2}\, H_3\wedge C_3-F_6\Bigr) \,, 
\label{eq:M5-action}
\end{align}
where $h_{\alpha\beta}\equiv G_{ij}\,\partial_\alpha X^i\,\partial_\beta X^j$, $H_3\equiv F_3-C_3$\,,
\begin{align}
 \theta^{\alpha}{}_\beta &\equiv \delta^\alpha_\beta + \frac{2}{3}\,\delta^{\alpha\alpha_1\alpha_2\alpha_3}_{\beta\beta_1\beta_2\beta_3}\,H_{\alpha_1\alpha_2\alpha_3}\,H^{\beta_1\beta_2\beta_3} 
\nn\\
 &= \Bigl[1+\frac{\tr(H^2)}{6}\Bigr]\,\delta^\alpha_\beta - \frac{1}{2}\,(H^2)^\alpha{}_\beta \,,
\end{align}
and $(H^2)^\alpha{}_\beta\equiv H^{\alpha \gamma_1\gamma_2}\,H_{\beta\gamma_1\gamma_2}$. 
Here and hereafter, indices are raised or lowered with $(h^{-1})^{\alpha\beta}$ or $h_{\alpha\beta}$\,. 
Using the equation of motion for $\gamma_{\alpha\beta}$, 
\begin{align}
 \sqrt{-\gamma}^{\,2}\,\bigl(\gamma^{-1}\,h\,\gamma^{-1}\,h\bigr)^\alpha{}_\beta = \sqrt{- h}^{\,2}\,\theta^\alpha{}_\beta \,,
\label{eq:EOM-M5-gamma}
\end{align}
we can eliminate $\gamma_{\alpha\beta}$ from the action,
\begin{align}
 S= - \int_\Sigma \frac{\rmd^6\sigma\sqrt{-h}\tr (\theta^{\frac{1}{2}})}{6}  + \int_\Sigma\Bigl(C_6-\frac{1}{2}\,H_3\wedge C_3-F_6\Bigr)\,. 
\nn
\end{align}
Now, if we consider the quadratic weak-field approximation in $H_3$, this action becomes
\begin{align}
 S &\sim -\int_\Sigma\rmd^6\sigma \sqrt{-h} 
         + \frac{1}{4}\int_\Sigma H_3\wedge *_h H_3 
\nn\\
   &\quad +\int_\Sigma \Bigl(C_6-\frac{1}{2}\, H_3\wedge C_3\Bigr) - \int_{\partial\Sigma} A_5 \,,
\end{align}
where the Hodge star $*_h$ is taken with respect to the induced metric $h_{\alpha\beta}$\,. 
This is essentially the same as the M5-brane action given in \cite{Bergshoeff:1996ev}. 
The equation of motion for $A_2$ gives $\rmd (*_h H_3+H_3)=0$, and it is consistent with the linearized (anti-)self-duality relation, $H_3=-*_h H_3$\,. 
At the nonlinear level, the equation of motion for $A_2$ obtained from \eqref{eq:M5-action}, together with \eqref{eq:EOM-M5-gamma}, is consistent with the relation $\cC_{[\alpha_1}{}^\alpha \, H_{\alpha_2\alpha_3]\alpha} = -(*_h H_3)_{\alpha_1\alpha_2\alpha_3}$, with
\begin{align}
 \cC_\alpha{}^\beta \equiv \frac{\tr(\theta^{-\frac{1}{2}})}{3}\,\delta_\alpha^\beta - (\theta^{-\frac{1}{2}})_\alpha{}^\beta \,.
\end{align}
The known nonlinear self-duality relation \cite{Howe:1996yn,Howe:1997fb,Sezgin:1998tm} has the form $C_{[\alpha_1}{}^\alpha \, H_{\alpha_2\alpha_3]\alpha} = -(*_h H_3)_{\alpha_1\alpha_2\alpha_3}$, with
\begin{align}
 C_\alpha{}^\beta = K^{-1}\,\Bigl\{\Bigl[1+\frac{1}{12}\,\tr(H^2)\Bigr]\,\delta_\alpha^\beta - \frac{1}{4}\,(H^2)_\alpha{}^\beta \Bigr\} \,,
\label{eq:open-membrane-metric}
\end{align}
where $K\equiv \sqrt{1+[\tr(H^2)/24]}$\,. 
Although $C_\alpha{}^\beta$ appears to be different from our matrix $\cC_\alpha{}^\beta$, we can show that they are the same matrix using the identity $(H^2H^2)_\alpha{}^\beta=(2/3) \tr(H^2)\,[\mathbf{1}+(1/2)\,H^2]_\alpha{}^\beta$ \cite{VanderSchaar:2001ay}, obtained from the self-duality relation for a flat $h_{\alpha\beta}$\,. 
Moreover, at least when the target space is flat, using the self-duality relation, we can easily show that the equation of motion for $X^i$ also has the same form as the known equation $\partial_\alpha\bigl(\sqrt{-h}\, C^{\alpha\beta}\,\partial_\beta X^j\bigr) = 0$ \cite{Howe:1997fb,Gibbons:2000ck}. 
We thus expect our theory to be equivalent to the conventional theory even at the nonlinear level, although the action apparently looks different from known ones: e.g., the Pasti-Sorokin-Tonin action \cite{Pasti:1997gx}. 

It is also interesting to note that the so-called open membrane metric $C^{\alpha\beta}$ \cite{Bergshoeff:2000jn} can be shown to be equal to $\gamma^{\alpha\beta}$ using the above relations. 

%\section{Actions for exotic branes}
{{\it Actions for exotic branes}.---}Each auxiliary field $\cP_I$ corresponds to a conventional brane in string or M-theory, and the choice of $\Omega_{p+1}$ considered in this Letter gives the worldvolume theory of the conventional branes only. 
In constant backgrounds where $\cM_{IJ}$ is independent of $x^i$, there is no reason to stick to the coordinates $X^i$ in constructing $\Omega_{p+1}$. 
In that case, we can instead use $\cQ_I \equiv \cM_{IJ}\,\cP^J$ as the fundamental fields and define the dual coordinates $\tilde{X}_i$ such that $\cQ_i = \rmd \tilde{X}_i$\,. 
Namely, we can rewrite our action in terms of $\cQ_I$ as
\begin{align}
 S = \frac{1}{p+1}\biggl[\,\frac{1}{2} \int_\Sigma (\cM^{-1})^{IJ}\,\cQ_I\wedge *\,\cQ_J - \int_\Sigma \Omega_{p+1}\,\biggr]\,.
\end{align}
In DFT, we can parametrize $(\cM^{-1})^{IJ}$ in the same way as \eqref{eq:DFT-gen-met} by using the non-geometric potential $\beta^{ij}$,
\begin{align}
 (\cM^{-1})^{IJ} =\begin{pmatrix} \mathbf{1} & \beta \cr \mathbf{0} & \mathbf{1}
 \end{pmatrix}\begin{pmatrix} \tilde{g}^{-1} & \mathbf{0} \cr \mathbf{0} & \tilde{g} 
 \end{pmatrix}\begin{pmatrix} \mathbf{1} & \mathbf{0} \cr -\beta & \mathbf{1}
 \end{pmatrix} \,. 
\end{align}
This is also the case in EFT \cite{LRS}, and we can generally construct a worldvolume theory for an exotic brane that electrically couples to a non-geometric potential (such as $\beta^{ij}$) by choosing $\Omega_{p+1}$ in the same way as we explained. 
However, this is not enough to obtain the worldvolume theories of all exotic branes. 
For example, the famous exotic $5^2_2$-brane and the $1^6_4$-brane magnetically couple to $\beta^{ij}$ and $\beta^{i_1\cdots i_6}$, respectively. 
In fact, they electrically couple to a certain mixed symmetry tensor $\beta_{i_1\cdots i_8,j_1j_2}$ or $\beta_{i_1\cdots i_8,j_1\cdots j_6}$ \cite{Sakatani:2014hba}, but these are not contained in $(\cM^{-1})^{IJ}$ for $d\leq 7$. 
Therefore, in order to describe all of the exotic branes, we need to construct $\Omega_{p+1}$ such that the magnetic couplings can be described or find a parametrization of $(\cM^{-1})^{IJ}$ in terms of the mixed-symmetry tensors. 

%\section{Conclusion and Outlook}
{{\it Conclusion and Outlook}.---}We proposed a bosonic action for a single brane from the perspective of extended field theories. 
Once a generalized vector $\cP^I$ and a parametrization of the generalized metric $\cM_{IJ}$ are given, we can automatically write down the worldvolume action for a $p$-brane that electrically couples to a $(p+1)$-form potential contained in $\cM_{IJ}$. 
As demonstrations, we showed that the known actions for a string, a membrane, and an M5-brane can be reproduced for the $E_{6(6)}$ case. 
By further considering the higher exceptional groups, $E_{7(7)}$ and $E_{8(8)}$, we may also reproduce actions for higher dimensional branes, and it is important to check whether the known actions, such as the action of the Kaluza-Klein monopole, can be reproduced correctly. 

In EFT, considering two parametrizations for $\cM_{IJ}$, we can derive both the 11-dimensional and the type IIB supergravity from a single EFT action \cite{Blair:2013gqa} (the explicit type IIB parametrization for $E_{d(d)}$ EFT with $d\leq 7$ is given in \cite{LRS}). 
We can apply our formulation also to the type IIB case to obtain worldvolume theories for various branes. 
Because of the success in the reproduction of M-theory brane actions, we expect that we can also reproduce the actions for branes in the type IIB string theory. 

From the perspective of \cite{Hull:2004in,Asakawa:2012px}, any D$p$-brane is a single ten-dimensional object in the doubled spacetime, and the value $p$ can be changed by duality rotations. 
As an extension of this idea, it is interesting to investigate a certain $\Omega$ that transforms covariantly under the duality transformations. 
Furthermore, the non-Abelian and supersymmetric extensions should be also studied. 

\begin{acknowledgments}
%{{\it Acknowledgments}.---}
We wish to thank Jeong-Hyuck Park, Soo-Jong Rey, and Satoshi Watamura for the helpful discussions. 
\end{acknowledgments}


\appendix

\section{Conventions}

We use the conventions, 
\begin{align}
\begin{split}
 &\varepsilon^{0\cdots p}=\frac{1}{\sqrt{-\gamma}}\,,\quad 
 \varepsilon_{0\cdots p}=-\sqrt{-\gamma} \,,
\\
 &\rmd^{p+1}\sigma = \rmd\sigma^0\wedge\cdots\wedge\rmd\sigma^p \,,
\\
 &(* w_q)_{\alpha_1\cdots\alpha_{p+1-q}} =\frac{1}{q!}\,\varepsilon^{\beta_1\cdots\beta_q}{}_{\alpha_1\cdots\alpha_{p+1-q}}\,w_{\beta_1\cdots\beta_q} \,,
\end{split}
\end{align}
for the epsilon tensor, the volume form, and the Hodge star operator. 
The eleven-dimensional supergravity fields are defined such that the field strengths have the form,
\begin{align}
 \mathcal{G}_4 \equiv \rmd C_4 \,,\quad 
 \mathcal{G}_7 \equiv \rmd C_6 + \frac{1}{2}\,C_3\wedge \rmd C_3 \,. 
\end{align}
Under a generalized diffeomorphism in $E_{6(6)}$ EFT with a gauge parameter, $V^I=(v^i,\,\tilde{v}_{i_1i_2}/\sqrt{2},\,\tilde{v}_{i_1\cdots i_5}/\sqrt{5!})$, the form fields transform as
\begin{align}
\begin{split}
 \delta_V C_3 &= \pounds_v C_3 + \rmd \tilde{v}_2\,,
\\
 \delta_V C_6 &= \pounds_v C_6 + \rmd \tilde{v}_5 - \frac{1}{2}\,C_3\wedge \rmd \tilde{v}_2 \,.
\end{split}
\end{align}
The closed $(q+1)$-forms, that specifies the embedding of the six-torus, also transform in the same manner, 
\begin{align}
\begin{split}
 \delta_V c_3 &= \pounds_v c_3 + \rmd \tilde{v}_2
 = \rmd \bigl(\tilde{v}_2 + \iota_v c_3\bigr)\,,
\\
 \delta_V c_6 &= \pounds_v c_6 + \rmd \tilde{v}_5 - \frac{1}{2}\,c_3\wedge \rmd \tilde{v}_2 
\\
 &= \rmd \Bigl(\tilde{v}_5 + \frac{1}{2}\,c_3\wedge \tilde{v}_2 + \iota_v c_6\Bigr) 
\,,
\end{split}
\end{align}
where we used the closedness. 
Noting that the field strengths $\{F_{q+1}\}$ transform in the same manner as the pull-back of $\{c_{q+1}\}$, we can summarize the transformation laws for $X^i(\sigma)$ and the gauge fields $A_q(\sigma)$ under the generalized diffeomorphism as
\begin{align}
\begin{split}
 \delta_V X^i &= v^i \,,
\\
 \delta_V A_2 &= \tilde{v}_2 + \iota_v F_3 \,,
\\
 \delta_V A_5 &= \tilde{v}_5 + \frac{1}{2}\,F_3\wedge \tilde{v}_2 + \iota_v F_6 \,.
\end{split}
\end{align}
From these, we can easily show the invariance of the combination, $\int_\Sigma\bigl[C_6-(1/2)\,H_3\wedge C_3-F_6\bigr]$\,. 

For a string in the doubled spacetime, the transformation laws under a generalized diffeomorphism become
\begin{align}
 \delta_V X^i = v^i \,, \quad 
 \delta_V A_1 = \tilde{v}_1 + \iota_v F_2 \,. 
\end{align}
Note that a similar expression (in the static gauge) is obtained in the construction of the D-brane worldvolume theory, Eq.~(4.7) in \cite{Asakawa:2012px}. 

\end{document}